\documentclass[11pt,aps,prd,preprint,superscriptaddress]{revtex4-1}
\usepackage{amsmath}
\usepackage[dvips]{graphicx}
\usepackage[english]{babel}
\begin{document}
\title{The generalized second law in the emergent universe}
\author{Sergio del Campo\footnote{E-mail: sdelcamp@ucv.cl}}
\affiliation{Instituto de F\'{\i}sica, Pontificia Universidad
Cat\'{o}lica de Valpara\'{\i}so, Av. Universidad 330, Campus
Curauma, Valpara\'{\i}so, Chile}
\author{Ram\'{o}n Herrera\footnote{E-mail: ramon.herrera@ucv.cl}}
\affiliation{Instituto de F\'{\i}sica, Pontificia Universidad
Cat\'{o}lica de Valpara\'{\i}so, Av. Universidad 330, Campus
Curauma, Valpara\'{\i}so, Chile}
\author{Diego Pav\'{o}n\footnote{E-mail: diego.pavon@uab.es}}
\affiliation{Departamento de F\'{\i}sica, Facultad de Ciencias,
Universidad Aut\'{o}noma de Barcelona, 08193 Bellaterra
(Barcelona), Spain}
\begin{abstract}
This paper studies whether the generalized second law of
thermodynamics is fulfilled in the transition from a generic
initial Einstein static phase to the inflationary phase, with
constant Hubble rate, and from the end of the latter to the
conventional  era of thermal radiation dominated expansion. As it
turns out, the said law is satisfied provided the radiation
component does not  largely contribute to the total energy of the
static phase.
\end{abstract}

\maketitle

\section{Introduction}
The highly successful hot big-bang  model presents the
uncomfortable feature of beginning the cosmic expansion out of an
initial singularity, a state of infinity energy density and
pressure where the laws of physics break down. To evade this
serious shortcoming several solutions have been proposed. These
include an ever bouncing universe (a universe that undergoes
infinite cycles of expanding-contracting phases), and the
so-called emergent universe.

\begin{figure}[htb]
\centering
\includegraphics[width=12cm]{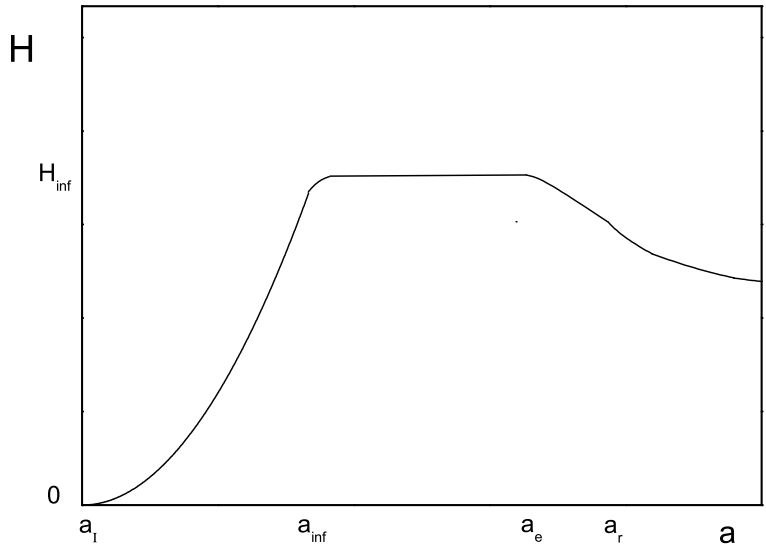}
\caption{Schematic evolution of the Hubble function from the
Einstein static era to the thermal radiation era. Here $a_{inf}$
and $a_{e}$ stand for the scale factor at the beginning and end of
exponential inflation, respectively; $a_{r}$ denotes the scale
factor at some generic point at the radiation dominated expansion
era where $w = \tilde{w}_{\gamma}$.} \label{fig:H(a)}
\end{figure}

The latter scenario replaces the initial singularity by an
Einstein static phase in which the scale factor of the
Friedmann-Robertson-Walker (FRW) metric does not vanish and,
accordingly, the energy density, pressure and so on do not
diverge. In the usual description the Universe starts expanding
from the said phase, smoothly joins a stage of exponential
inflation  followed by standard reheating to subsequently approach
the classical thermal radiation dominated era of the conventional
big bang cosmology \cite{cqg1-george,cqg2-george}. Figure
\ref{fig:H(a)} summarizes this evolution.

This Letter studies which constraints (if any) the generalized
second law (GSL) of thermodynamics imposes on the two intermediate
phases, i.e., from the static phase to exponential inflation and
from the reheating to thermal radiation domination (in the static,
inflationary, and thermal radiation dominated phases the GSL is
trivially fulfilled). According to the GSL the entropy of the
horizon plus the entropy of the matter and fields within the
horizon cannot decrease \cite{jakob,gsl,brustein}. In the present
case neither the particle horizon nor the event horizon exist in
the static phase; only the apparent horizon  does exist in all the
phases dealt with here, therefore this is the one to be
considered. Moreover, it has been reasoned that the latter horizon
is the appropriate one in dealing with the second law of
thermodynamics \cite{bye}.

The apparent horizon in FRW universes is defined as the marginally
trapped surface with vanishing expansion \cite{bak-rey};
accordingly, its radius is given by $\tilde{r}_{A} =
1/\sqrt{H^{2}\, + \, k \, a^{-2}}$, where $H = \dot{a}/a$ denotes
the Hubble function and $a$ the scale factor of the FRW metric
(for reasons given below we just consider $k = +1$).

As is widely admitted, the entropy of the apparent horizon is not
simply proportional to the horizon area but it includes quantum
modifications,
\begin{equation}
S_{{\cal A}} = k_{B} \, \left[\frac{{\cal A}}{4\, \ell^{2}_{pl}}
\, + \, \alpha \, \ln \frac{{\cal A}}{4\, \ell^{2}_{pl}}\right] \,
. \label{eq:S(A)1}
\end{equation}
Here, $k_{B}$ is Boltzmann's constant, $\ell_{pl}$ Planck's
length, and $\alpha \, $ an $\, {\cal O}(1)$ constant. The first
term is the classical one and varies directly as the area  of the
apparent horizon, ${\cal A} = 4 \pi \tilde{r}_{A}^{2}$, while the
second one arises from quantum corrections
\cite{qcorrect1,qcorrect2,qcorrect3,plb-nd}. Therefore, in those
situations in which the entropy inside the horizon can be ignored
the second law of thermodynamics simply imposes that the
constraint
\begin{equation}
S'_{\cal A} = k_{B} \, \left[ \frac{1}{4 \, \ell^{2}_{pl}}\, + \,
\frac{\alpha}{{\cal A}}\right] \,  {\cal A}' \geq 0\,
\label{eq:1stderiv}
\end{equation}
must be satisfied at all times. Here ${\cal A}' = -  ({\cal
A}^{2}/2\pi) \, (HH'\, -\, k\, a^{-3})\, $  and the prime means
derivative with respect to the scale factor.

Section II briefly summarizes the main features of the emergent
scenario of Refs. \cite{cqg1-george} and \cite{cqg2-george}.
Section III examines whether the GSL is satisfied in the
transitions from the static phase to exponential inflationary
phase and from the end of the latter to thee xpansion era
dominated by thermal radiation. Finally, section IV summarizes our
findings.

\section{The emergent scenario}
In this scenario the Universe begins expanding from a finite
initial size, say $a_{I}$, thus avoiding the  singularity that
lurks at the starting point of the standard big-bang model. Since
one is free to choose $a_{I} \gg \ell_{pl}$ the quantum gravity
era is also avoided. No exotic energy is needed. It suffices, e.g.
ordinary matter (subscript $m$), radiation (subscript $\gamma$),
and a quintessence scalar field $\phi$  with potential $V(\phi)$
obeying Klein-Gordon equation -actually, in the static phase
matter and radiation are not strictly necessary but we know of no
obvious reason to exclude them. The total energy density in this
phase is simply $\rho_{m,I}\, + \, \rho_{\gamma,I} \, +
(1/2)\dot{\phi}^{2}_{I} \, + V_{I} = \frac{3k}{8 \pi G
a_{I}^{2}}$, which implies that this universe must have closed
spatial sections (this is why we set $k = +1$).

To implement the scenario the potential must be asymptotically
flat in the infinite past,
\begin{equation}
V(\phi) \rightarrow V_{I} \quad {\rm as} \quad \phi \rightarrow
-\infty\, , \quad t \rightarrow - \infty \, ,
\label{eq:Vflat}
\end{equation}
and fall toward a minimum at some finite value. As a consequence,
the quintessence field rolls down from the static state at $-
\infty$ and the potential slowly decreases from its initial value,
$V_{I}$, in the infinite past. To have acceleration the inequality
$V(\phi) - \dot{\phi}^{2} > 0$  must be fulfilled. Since $V(\phi)$
decreases and $\dot{\phi}^{2}$ augments, at some time, say $t=
t_{e}$, inflation terminates, then $\phi$ oscillates about the
minimum and reheating takes place, the latter followed by the
radiation dominated era -see references \cite{cqg1-george} and
\cite{cqg2-george} for details.

\section{The GSL at the transition phases}
\subsection{Transition from the static to the inflationary phase}
This period corresponds to the scale factor interval $a_{I} < a <
a_{inf}$ in Fig. \ref{fig:H(a)}.

During the static phase, $H_{I} = 0$, of this closed universe the
scale factor, $a = a_{I}$, remains constant and so the total
energy density, $\rho_{I} = 3 k/(8\pi G a_{I}^{2}) \, $
\cite{cqg1-george}. Accordingly, the horizon, whose area reduces
to $ {\cal A}_{I} = 4\pi \, a_{I}^{2}$, exists just because the
spatial curvature is positive.

In this transition $H'> 0$ but the sign of $ {\cal A}' $ rests
also on the curvature and therefore on the relationship between
$H$ and the scale factor, i.e., it crucially depends on the
specific model. As in the emergent scenario of \cite{cqg1-george}
and \cite{cqg2-george}, if the model conforms general relativity,
we can write
\begin{equation}
H \, H' \, - \, \frac{k}{a^{3}} = -4 \pi \, G\, \frac{\rho \, + \,
p}{a} \, , \label{eq:Hprimegr}
\end{equation}
and
\begin{equation}
{\cal A}' = 2 G \, {\cal A}^{2} \; \frac{\rho \, + \, p}{a} \, ,
\label{eq:Aprimegr}
\end{equation}
whence the horizon area augments only when the overall gravity
source respects the dominant energy condition. The compatibility
of the latter  with the increase of the Hubble factor might look
counter-intuitive. Nevertheless,  it can be realized -for
instance- in universes whose expansion is dominated by matter
and/or radiation and a quintessence scalar field. It is worthy of
note that the positive spatial curvature makes possible both the
Einstein static phase and the increase of the horizon area during
this transition phase.

Still, $S'_{\cal A}$ could be negative if the square parenthesis
on the right hand side of (\ref{eq:1stderiv}) were negative.
However, for this to occur one should have $ \alpha < - {\cal
A}/(4 \ell_{pl}^{2})$. Since it is rather reasonable to expect $
\ell^{2}_{pl} \ll {\cal A}$ at all times, for the parenthesis to
be negative  $\alpha$ should take a very large negative value,
something highly disfavored \cite{cqg-hod}.

At this point one may wonder whether the entropy of the matter
and/or radiation in the horizon volume may increase or decrease
(we do not consider the entropy of the scalar field for we assume
that the latter is in a pure quantum state). We first consider
pressureless matter. At any given instant the entropy of dust
inside the horizon  obeys $S_{m} = k_{B}\, (4\pi/3) \,
\tilde{r}_{A}^{3} \,  n $, where $n = 3N/ (4 \pi \,a_{I}^{3})$ is
the number density of dust particles and $N$ the total number
particles in the static phase of radius $a_{I}$. In consequence,
\begin{equation}
S'_{m} = - 3 k_{B} \, \frac{N}{a_{I}^{3}}\,\tilde{r}_{A}^{5} \,
\left(H \, H' \, - \, \frac{k}{a^{3}}\right).
\label{eq:Smatterderiv}
\end{equation}
Again, if the model is governed by general relativity and complies
with the dominant energy condition (i.e., $\rho \, + \, p > 0$),
Eq. (\ref{eq:Hprimegr}) ensures that $S'_{m}$ will be
positive-definite.

For radiation (as for any fluid possessing pressure) the entropy
obeys Gibbs equation \cite{callen}. In our case,
\begin{eqnarray}
T_{\gamma} \, S'_{\gamma} = \frac{d}{da} \left(\frac{4 \pi}{3} \,
\tilde{r}_{A}^{3} \, \rho_{\gamma}\right) \, + \, w_{\gamma} \,
\rho_{\gamma} \frac{d}{da}\left(\frac{4 \pi}{3}\,
\tilde{r}_{A}^{3}\right) &= & 2 \pi \, (1\, + \, w_{\gamma}) (1 \,
+ \, 3w) \tilde{r}_{A}^{3} \frac{\rho_{\gamma}}{a} \, .
\label{eq:gibbs1}
\end{eqnarray}
In arriving to the second equality we have used $\rho'_{\gamma}= -
3 (1+w_{\gamma})\, \rho_{\gamma}/a \, $ and Eq.
(\ref{eq:Hprimegr}). Because of the long duration of the static
phase we can safely assume that the radiation is thermalized
(i.e., it presents a black-body distribution) thereby we can set
$w_{\gamma} = 1/3$. Since the expansion is accelerated the
equation of state of the overall fluid (including the scalar
field), $w = p/\rho$, must fulfill $1 \, + \, 3w < 0$ whence
$S'_{\gamma}$ results negative. In consequence, for the GSL to be
satisfied the radiation component must not  largely contribute to
the total energy density. To be more specific, the GSL condition
$S'_{{\cal A}} \, + \, S'_{m}\, + \, S'_{\gamma} \geq 0$
translates into the constraint
\begin{equation}
\frac{\rho_{\gamma}}{\rho} \leq \frac{3}{4}\, \frac{k_{B} G \pi
(1\, + \, w) T_{\gamma I} \, a_{I} \, \tilde{r}_{A}
\left[\frac{4}{\ell^{2}_{pl}} \, + \, 6 \frac{N}{a_{I}^{3}}\,
\tilde{r}_{A}\right]}{|1\, + \, 3w|\, a}\, . \label{constraint1}
\end{equation}
In writing last equation we have neglected the second term in the
square parenthesis in Eq. (\ref{eq:1stderiv}) and made use of the
evolution law for the radiation temperature $T_{\gamma} =
T_{\gamma I} (a_{I}/a)$, where $T_{\gamma I}$ stands for its value
at the Einstein static phase. Since the ratio $\rho_{\gamma}/\rho
\propto a^{3w}$ (with $w <0$) decreases with expansion, the
maximum value of the right hand side of (\ref{constraint1}) occurs
when $a =a_{I}$. This implies the upper bound
\begin{equation}
\frac{\rho_{\gamma}}{\rho} \leq \frac{3}{4}\, \frac{k_{B} G \pi
(1\, + \, w) T_{\gamma I} \, a_{I} \left[\frac{4}{\ell^{2}_{pl}}
\, + \, 6 \frac{N}{a_{I}^{2}}\right]}{|1\, + \, 3w|}\, ,
\label{constraint2}
\end{equation}
which must be respected if the GSL is to be fulfilled.

\subsection{Transition between the exponential inflation regime and the thermal
radiation dominated phase}
This period starts at $a = a_{e}$ and terminates when the products
of the inflaton decay get fully thermalized.

The exponential inflationary period is characterized by $H =$
constant. At the end of it -according to cold inflation- a very
fast phase of reheating takes place. In the course of the latter
all the energy of the inflaton field is very quickly converted
into a mixture of radiation and relativistic particles that begins
dominating the expansion. Automatically, the strong energy
condition, $\rho \, + \, p > 0$, is met and the Universe starts
decelerating with $H' <0$. However, since the mixture of radiation
and particles produced by the decay of the inflaton takes some
time (by all accounts much larger than the exponential inflation
stage) in thermalize, the conventional thermal radiation dominated
era (in which $w_{\gamma} = 1/3$) cannot commence right away, it
has to wait. Here we analyze whether the GSL is fulfilled during
the period connecting the reheating phase with the conventional
thermal dominated phase.

Again ${\cal A}'$ is given by Eq. (\ref{eq:Aprimegr}) where now
$\rho = \rho_{\gamma} $  and $p = p_{\gamma}$, whence  $S'_{{\cal
A}} > 0$. Likewise the entropy variation of the mixture of
relativistic particles and radiation (of total energy
$\rho_{\gamma}$)  is given by the right hand side of  Eq.
(\ref{eq:gibbs1}) with  $w$ and $w_{\gamma}$ replaced by
$\tilde{w}$, the latter being positive-definite (the tilde is to
remind us that the mixture is not thermalized though, obviously,
$\tilde{w}_{\gamma} \rightarrow w_{\gamma} = 1/3$ as the
thermalization  proceeds). Thus, the GSL is guaranteed in this
transition.

We have not considered dissipative processes like the decay of
heavy (but still relativistic) particles into lighter ones and
bulk viscosities. However, these processes being irreversible in
nature necessarily augments the phase space and, accordingly, the
entropy within the horizon.

\section{Concluding remarks}
A successful cosmological scenario is expected to fulfill the laws
of thermodynamics, in particular the GSL. In this paper we found
that the condition for the emergent scenario of Refs.
\cite{cqg1-george} and \cite{cqg2-george} to comply with the GSL
is that the radiation component do not contribute largely to the
total energy density of the static phase. More precisely, that the
bound (\ref{constraint2}) be met at the commencement of the
transition from the static phase to the period of exponential
inflation where $H =$ constant. Most emergent scenarios are
expected to easily fulfill this condition since for the Universe
to transit from the static phase ($H = H_{I} = 0$) to the said
period its expansion must be dominated by some energy component
that violates the strong energy condition -as for instance a
quintessence scalar field. It should be noted that although
neither radiation nor pressureless matter are necessary
ingredients of the energy budget in the Einstein static phase it
is quite natural to consider them. Clearly if radiation were
absent, the GSL would  be more readily satisfied.

We have ignored the exponential inflationary phase (i.e., the
interval running from $a_{inf}$ to $a_{e}$) because solely thermal
radiation could contribute negatively  to the total $S'$ in this
period. Such contribution, being proportional to $\rho_{\gamma}$
(see Eq. (\ref{eq:gibbs1})), is -in any case- vanishing since, due
to the huge expansion from $a_{i}$ to $a_{e}$ in the emergent
model under consideration, this component gets practically
redshifted away before the onset of the said period.

While we focused on one specific scenario it encapsulates the main
features any emergent model must possess. This is why we believe
the overall result of this research  should bear a wide generality
and, in any case, it may serve as a springboard to the study of
the GSL in more sophisticated emergent scenarios.

\acknowledgments{This work was supported from ``Comisi\'{o}n
Nacional de Ciencias y Tecnolog\'{\i}a" (Chile) through the
FONDECYT Grant No. 1110230 (SdC) and  No. 1090613 (RH and SdC).
D.P. acknowledges ``FONDECYT-Concurso incentivo a la cooperi\'{o}n
internacional" No. 7080212, and is grateful to the ``Instituto de
F\'{\i}sica" for warm hospitality; also D.P. research was
partially supported by the ``Ministerio Espa\~{n}ol de
Educaci\'{o}n y Ciencia" under Grant No. FIS2009-13370-C02-01, and
by the ``Direcci\'{o} de Recerca de la Generalitat" under Grant
No. 2009SGR-00164.}

\end{document}